\documentclass[12pt]{article}
\usepackage{arxiv}
\usepackage[utf8]{inputenc} 
\usepackage[T1]{fontenc}    
\usepackage{hyperref}       
\usepackage{url}            
\usepackage{booktabs}       
\usepackage{amsfonts}       
\usepackage{nicefrac}       
\usepackage{microtype}      
\usepackage{lipsum}
\usepackage{graphicx}
\usepackage{amsmath}
\usepackage{amsthm}
\usepackage{tikz}
\usepackage{booktabs}
\usepackage{amssymb}
\usepackage{multirow}
\usepackage{placeins}
\usepackage{color,soul}
\usepackage{xcolor}
\usepackage{subcaption}
\usepackage{enumitem}
\usepackage{tabularx, booktabs, multirow, siunitx}
\usepackage{comment}
\usepackage{siunitx}
\newcolumntype{E}{S[table-format=1.2]}          
\newcolumntype{C}{S[table-format={[2.2,\,2.2]}]} 

\newtheorem*{assumption*}{\assumptionnumber}
\providecommand{\assumptionnumber}{}

\makeatletter
 {%
  \renewcommand{\assumptionnumber}{Assumption $\mathcal{#1}$}%
  \begin{assumption*}%
  \def\assumpLetter{#1}
  \protected@edef\@currentlabel{$\mathcal{#1}$}%
  \renewcommand{\labelenumi}{$\mathcal{#1}$-\theenumi}%
  \renewcommand{\p@enumi}{$\mathcal{#1}$-} 
 }
 {%
  \end{assumption*}
 }
\makeatother

\usetikzlibrary{bayesnet, positioning, fit, calc}
\usepackage[style=authoryear, backend=biber]{biblatex}
\graphicspath{ {./images/} }

\title{Regression Analysis After Bipartite Bayesian Record Linkage}

\author{
    Xueyan Hu\\
    Department of Statistical Science\\
    Duke University\\
    Durham, NC 27705\\
    \texttt{xh149@duke.edu} \\
\And
    Jerome P. Reiter \\
    Department of Statistical Science\\ 
    Duke University\\
    Durham, NC 27705 \\
    \texttt{jreiter@duke.edu}
}

\begin{document}
\maketitle
\begin{abstract}
In many settings, a data curator links records from two files to produce datasets that are shared with secondary analysts. Analysts use the linked files to estimate models of interest, such as regressions.  Such two-stage approaches do not necessarily account for uncertainty in model parameters that results from uncertainty in the  linkages. Further, they do not leverage the relationships among the study variables in the two files to help determine the linkages. 
We propose a multiple imputation framework to address these shortcomings. First, we use a bipartite Bayesian record linkage model to generate multiple plausible linked datasets, disregarding the information in the study variables.  Second, we presume each linked file has a mixture of true links and false links. We estimate the mixture model using information from the study variables. Through simulation studies under a regression setting, we demonstrate that estimates of the regression model parameters can be more accurate than those based on an analogous two-stage approach. We illustrate the integrated approach using data from the Survey on Household Income and Wealth, examining a regression involving the persistence of income.
\end{abstract}


\section{Introduction}
Record linkage, the process of identifying records that refer to the same entity across multiple datasets, is an important component of data integration. With linked files, researchers can conduct more comprehensive investigations than from a single data source alone. Thus, it is frequently used in many fields including public health \parencite{jutte2011administrative, dzomba2025viral}, economics \parencite{abowd2019optimal, bailey2020well} , and policy research \parencite{ruggles2018historical, bailey2020well}. 

We consider settings where a data curator, such as a government agency or healthcare organization, seeks to link two files to create datasets that it shares with secondary analysts.  Such record linkage can be straightforward when the data curator has access to perfectly recorded, unique identifiers on both files. Often, however, such identifiers are not available, for example, because of privacy concerns or because they were never collected in one or both files in the first place. As a result, the data curator must link the records using variables present on both files, such as names, addresses, and birth dates. These variables can be prone to recording errors, variations, and missing values, creating uncertainty in the linkage process. Linking in imperfect variables inevitably results in imperfect linkages. In turn, these imperfect linkage affect the quality of downstream analyses \parencite{RN2, di2018adjusting, wang:slawski, chambers:dasilva}. Ideally, secondary analysts incorporate the uncertainty from imperfect linkage into their analyses. 

The literature on record linkage describes several approaches for incorporating imperfect linkages in downstream analyses, although in practice often this is not done \parencite{reiter:review}.  The most common approach is what we call the two-stage approach. These approaches begin with a linked file and apply corrections to parameter estimates to improve inferences  \parencite{scheuren1993regression, scheuren1997regression, lahiri2005regression, chambers2009regression, chambers2009inference, zhang2021linkage}. These methods are tailored to specific models and presume availability of quantities like the probabilities that any two records are a link, which may not be provided in the linked files made available to secondary analysts.
Another set of two-stage approaches relies on the expectation-maximization (EM) algorithm \parencite{chipperfield2011maximum, slawski2024general, hof2015mixture}.  These approaches operate on a single linked dataset  and do not necessarily account fully for the uncertainty introduced by the linkage process.

A second general approach is what we call integrated methods. In these approaches, the analyst simultaneously estimates the linkage model and the substantive analysis \parencite{dalzell:reiter, guha2022bayesian, shan2023bayesian, gutman2024multiple}.  For example, \textcite{tang2020bayesian} specify a Bayesian model in which the study variables  follow a linear regression for true links and a marginal distribution for false links, and the linkage follows the bipartite record linkage model of \textcite{sadinle2017bayesian}. Simultaneous linkage and inference for model parameters is conducted via Gibbs sampling. One advantage of integrated approaches is that they can utilize the information in study variables encoded by the regression to inform the linkages. For example, if two records in file $D_1$ are equally likely links for some record in file $D_2$, the integrated model can favor the candidate that best accords with the regression relationship.  In practice, however, these integrated techniques can be computationally expensive. Additionally, and particularly germane to our setting, integrated methods require access to all records in the files used for linkage. We seek methods that can be applied by secondary analysts who only have access to the linked data.

In this article, we propose an approach that seeks a compromise of the two-stage and integrated approaches.
First, we employ a Bayesian record linkage model \textcite{sadinle2017bayesian} to generate multiple plausible linked data files.  We then adapt a strategy developed by \textcite{slawski2024general} for the multiple plausible linked files, in which we conceive of each linked file as comprising a mixture of true and false links.  We estimate the mixture model in each file using an EM algorithm that incorporates the downstream regression model of interest.  Effectively, using the model offers the chance to downweight the contributions of apparent links that do not accord with the estimated regression model.
Finally, we combine the point and variance estimates of the regression parameters from the analysis of each plausibly linked file using multiple imputation inferences \parencite{rubin:1987}. 

We develop two strategies for implementing this approach, depending on whether or not the linked files include 
information about the quality of each linked record pair. 
We illustrate the performance of both methods
using simulation studies.
When the signal-to-noise ratio in the regression is not low e.g., $R^2 \geq 0.6$, the models can result in inferences that are  comparable to regression results using perfectly linked data. For scenarios with  low signal-to-noise ratios, e.g., $R^2 \leq 0.3$, we show that our approach needs some pre-determined correct links, which we call seeds, to give accurate results. 
In comparing the simulation results to those of a two-stage analysis that uses solely the plausibly linked files from the Bayesian record linkage, we find that the accuracy of our approach typically matches or exceeds the accuracy of the two-stage approach.

The remainder of this article is organized as follows. Section \ref{sec:background} reviews the methodological foundations underlying our approach. Section~\ref{sec:PLMIc} presents the proposed model incorporating linkage quality information and details its inference procedure. Section~\ref{sec:PLMI} describes a variant of this model that can be used without linkage quality metrics. Section \ref{sec:simu_study} presents results of simulation studies for both models.
Finally, Section \ref{sec:discussion} concludes with a discussion of implications and future directions.

\section{Background}\label{sec:background}

In Section \ref{sec:BRLmodel}, we review the bipartite  Bayesian record linkage model of \textcite{sadinle2017bayesian}.  In Section \ref{sec:slawski}, we review the post-linkage inference framework developed by \parencite{slawski2024general}, which we extend and modify in our methodology. In Section \ref{sec:multi_impu}, we review multiple imputation inferences \parencite{rubin2018multiple}. 

As context, we consider settings with two datasets $D_1$ and $D_2$ comprising $n_1$ and $n_2$ records, respectively.  Without loss of generality, we assume $n_1 \geq n_2$. Each dataset is assumed to be duplicate-free, meaning that no individual appears more than once within either $D_1$ or $D_2$. Both datasets contain $F$ common variables, referred to as {linking variables}, which serve as the basis for record linkage. Additionally, $D_1$ contains study variables of interest denoted by $\boldsymbol{X}$, and  $D_2$ contains study variables of interest denoted by $\boldsymbol{Y}$. Our primary goal is to conduct statistical inference on the regression of $\boldsymbol Y$  on   $\boldsymbol{X}$ while accounting for the uncertainty inherent in the record linkage process.  Here, we presume that $\boldsymbol Y$ includes a single variable and $\boldsymbol X$ includes $p$ variables.

\subsection{Bayesian Record Linkage Model}
\label{sec:BRLmodel}

For any pair of records $(i,j)$, where $i=1,\dots,n_1$ indexes a record in $D_1$ and $j = 1,\dots, n_2$ indexes a record in $D_2$, 
let $\boldsymbol{\gamma}_{ij} = (\gamma_{ij}^1,\dots,\gamma_{ij}^f, \dots, \gamma_{ij}^F)$ be $F$-dimensional integer-valued vector representing the similarity between record $i$ and record $j$, Here, 
each $\gamma_{ij}^f$ takes integer values in $\{1, \dots, L_f\}$, with larger values indicating greater similarity. We define each $\gamma_{ij}^f$ depending the type of information provided by linking variable $f$.  For instance, when linking variable $f$ is categorical such as a person's zip code, we may set $\gamma_{ij}^f=2$ when record $i$ and record $j$ have the exact same value of linking variable $f$, and $\gamma_{ij}^f=1$ otherwise. For string fields such as names and addresses, we first can compute string metrics such as Levenshtein edit distance \parencite{winkler1990string},  and subsequently discretize the distances into ordered levels based on predefined thresholds.  For example, when linking variable $f$ is a name, one sets $\gamma_{ij}^f=4$ when the last names match exactly, $\gamma_{ij}^f=3$ when the string metric is at or above 0.9 but below 1, $\gamma_{ij}^f=2$ when the string metric is at or above 0.75 but below 0.9, and $\gamma_{ij}^f=1$ when the string metric is below 0.75.  In general, record pairs where $\gamma_{ij}$ includes many high values, i.e., many similar fields, are more likely to be true links than record pairs where $\gamma_{ij}$ includes many low values.

For each record $j=1, \dots, n_2$ in $D_2$, let $Z_j$ indicate its linkage status.  Specifically, we set $Z_j=i$ when record $j$ in $D_2$ links to record $i$ in $D_1$.  We set $Z_j= n_1 + j$ when record $j$ does not have any link in $D_1$.  Let 
$\boldsymbol{Z} = (Z_1, \dots, Z_{n_2})$ to represent which records in $D_1$ are linked to records in $D_2.$ 

Let $\boldsymbol{\Gamma}_{ij} = (\Gamma_{ij}^1, \dots, \Gamma_{ij}^F)$ be a random variable representing possible realizations of $\boldsymbol{\gamma}_{ij}$.  For $f=1, \dots, F$, let  $\mu_{fl} = \Pr(\Gamma_{ij}^f = l \mid Z_j = i)$ and $\nu_{fl} = \Pr(\Gamma_{ij}^f = l \mid Z_j \neq i)$ denote the probabilities of observing agreement level $l$ in field $f$ for record pair $(i, j)$, conditional on whether the pair is a link or a non-link, respectively.
For $f=1, \dots, F$, let $\boldsymbol{\mu}_f = (\mu_{f1},\dots, \mu_{fL_f})$ and  $\boldsymbol{\nu}_f = (\nu_{f1},\dots, \nu_{fL_f})$.  Let $\boldsymbol{\mu} = (\boldsymbol{\mu}_1,\dots, \boldsymbol{\mu}_F)$ and $\boldsymbol{\nu} = (\boldsymbol{\nu}_1,\dots, \boldsymbol{\nu}_F)$.  Finally, let   $\boldsymbol{\Phi} = (\boldsymbol{\mu}, \boldsymbol{\nu})$.
We presume the record linkage model of \parencite{fellegi1969theory} and \parencite{sadinle2017bayesian}, who posit that 
\begin{equation}
    \boldsymbol{\Gamma}_{ij} \mid Z_j = i \overset{\text{iid}}{\sim} \mathcal{M}(\boldsymbol{\mu}),\quad
    \boldsymbol{\Gamma}_{ij} \mid Z_j = n_1 + j \overset{\text{iid}}{\sim} \mathcal{U}(\boldsymbol{\nu}).
\end{equation}
where $\mathcal{M}(\boldsymbol{\mu})$ and $\mathcal{U}(\boldsymbol{\nu})$ are models for the linked and non-linked record pairs with parameters $\boldsymbol{\mu}$ and $\boldsymbol{\nu}$. 
The model presumes that  (i) each  $\Gamma_{ij}^f$ where $f=1, \dots, F$ is independent and that (ii) each  
$\boldsymbol{\Gamma_{ij}}$ is conditionally independent given $\boldsymbol{Z}$. 
Hence, the model for the comparison vectors is 
\begin{equation}
    \Pr(\boldsymbol{\Gamma} = \boldsymbol{\gamma} \mid \boldsymbol{Z}, \boldsymbol{\Phi})
    = \prod_{i=1}^{n_1} \prod_{j=1}^{n_2} \prod_{f=1}^{F} \prod_{l=1}^{L_f} 
    \left( \mu_{fl}^{\mathbb{I}(Z_j = i)} \, \nu_{fl}^{\mathbb{I}(Z_j \neq i)} \right)^{\mathbb{I}(\gamma_{ij}^f = l)}.
    \label{eq: BRL_likelihood}
\end{equation}
The expression in \eqref{eq: BRL_likelihood} also serves as the likelihood function for the unknown parameters, $\mathcal{L}(\boldsymbol{Z}, \boldsymbol{\Phi}; \boldsymbol{\gamma})$.

To complete the specification of the Bayesian version of the model of \textcite{fellegi1969theory}, we need to set prior distributions on $\boldsymbol Z, \boldsymbol \Phi$.  We use the prior distributions specified by \textcite{sadinle2017bayesian}. For 
$\boldsymbol{\mu}_f$ and $\boldsymbol{\nu}_f$, we presume $\boldsymbol{\mu}_f\sim \text{Dirichlet}(\alpha_{f1},\dots, \alpha_{fL_f})$ and $\boldsymbol{\nu}_f\sim \text{Dirichlet}(\beta_{f1}, \dots, \beta_{fL_f})$. In the simulation studies of Section \ref{sec:simu_study}, we set all hyperparameters in these two prior distributions equal to one.  For the prior distribution of $\boldsymbol{Z}$, we first define an indicator variable whether record $j$ in $D_2$ has a link, $\mathbb{I}(Z_j \leq n_1) \overset{\text{iid}}{\sim}\text{Bernoulli}(\pi)$, with $\pi \sim \text{Beta}(\alpha_\pi, \beta_\pi)$.  The total number of links in $\boldsymbol{Z}$ is then $n_{12}(\boldsymbol{Z}) = \sum_{j=1}^{n_2}\mathbb{I}(Z_j \leq n_1)$. Given $n_{12}(\boldsymbol{Z})$, \textcite{sadinle2017bayesian} presumes that each feasible $\boldsymbol{Z}$ is equally likely.  Therefore, the prior of $\boldsymbol Z$ is given by 
\begin{equation}
    p(\boldsymbol{Z} \mid \alpha_\pi, \beta_\pi) = \frac{(n_1 - n_{12}(\boldsymbol{Z}))!}{n_1 !} \frac{B(n_{12}(\boldsymbol{Z}) + \alpha_\pi, n_2 - n_{12}(\boldsymbol{Z}) + \beta_\pi)}{B(\alpha_\pi, \beta_\pi)},
\end{equation}
where $B(\cdot,\cdot)$ is the Beta function.

We now describe the Gibbs sampler proposed by \textcite{sadinle2017bayesian} for the Bayesian record linkage model. Following the suggestion in \textcite{tang2020bayesian}, we initialize $\boldsymbol{Z}$ by running a non-Bayesian, probabilistic record linkage \parencite{fellegi1969theory}. We initialize each $\boldsymbol{\mu}_f$ and $\boldsymbol{\nu}_f$ with a random sample from their prior distributions. Given a current draw of $(\boldsymbol{\mu}^{(m)}, \boldsymbol{\nu}^{(m)}, \boldsymbol{Z}^{(m)})$ at iteration $m$, we update the parameters by sampling  $(\boldsymbol{\mu}^{(m+1)}, \boldsymbol{\nu}^{(m+1)}, \boldsymbol{Z}^{(m+1)})$ via steps {\bfseries S.1} and {\bfseries S.2}.

\paragraph{S.1} Sample the entries of $\boldsymbol{Z}^{(m+1)}$ sequentially. Having sampled the first $j-1$ entries of $\boldsymbol{Z}^{(m+1)}$, we define 
\begin{equation}
\boldsymbol{Z}^{(m+\frac{j-1}{n_2})}_{-j} = \left(Z_1^{(m+1)}, \ldots, Z_{j-1}^{(m+1)}, Z_j^{(m)}, \ldots, Z_{n_2}^{(m)}\right).
\end{equation}

We sample a new label $Z_j^{(m+1)}$, with the probability of selecting label $q \in \{1, \ldots, n_1, n_1 + j\}$ given by 
$p_{qj}\left( \boldsymbol{Z}^{(m+\frac{j-1}{n_2})}_{-j} \mid \boldsymbol{\Phi}^{(m+1)} \right).$
This can be expressed for generic $\boldsymbol{Z}_{-j}$ and $\boldsymbol{\Phi}$ as
\begin{equation}
p_{qj}(\boldsymbol{Z}_{-j} \mid \boldsymbol{\Phi}) \propto 
\begin{cases}
\exp(c_{qj}^{(m)}) \, \mathbb{I}(Z_{j'} \neq q ,\, \forall j' \neq j), 
    & \text{if } q \leq n_1, \\[1.5ex]
\left( n_1 - n_{12}(\boldsymbol{Z}_{-j}) \right) 
\dfrac{n_2 - n_{12}(\boldsymbol{Z}_{-j}) - 1 + \beta_\pi}
      {n_{12}(\boldsymbol{Z}_{-j}) + \alpha_\pi}, 
    & \text{if } q = n_1 + j
\end{cases}
\end{equation}
where $c_{qj}^{(m)}$ is the confidence measure for pair $[q,j]$,
\begin{equation}\label{eq:confidence_measure}
     c_{qj}^{(m)} = \sum_{f=1}^F \sum_{l=1}^{L_f} \mathbb{I}(\gamma_{qj}^f=l)\log(\frac{\mu_{fl}^{(m)}}{\nu_{fl}^{(m)}}).
\end{equation}

When we know for certain that record $j$ in $D_2$ matches to record $i$ in $D_1$, we set $Z_j=i$ directly.  

\paragraph{S.2} Sample $\boldsymbol{\mu}^{(m+1)}$ and $\boldsymbol{\nu}^{(m+1)}$. For $f = 1,\dots, F$, sample

\begin{equation}
    \begin{aligned}
        \boldsymbol{\mu}_f^{(m+1)} \mid \boldsymbol{\gamma}, \boldsymbol{Z}^{(m+1)} 
        &\sim \text{Dirichlet}\left(
            \alpha_{f0}(\boldsymbol{Z}^{(m+1)}) + \alpha_{f0},\,
            \dots,\,
            \alpha_{fL_f}(\boldsymbol{Z}^{(m+1)}) + \alpha_{fL_f}
        \right)\\
        \boldsymbol{\nu}_f^{(m+1)} \mid \boldsymbol{\gamma}, \boldsymbol{Z}^{(m+1)} 
        &\sim \text{Dirichlet}\left(
            \beta_{f0}(\boldsymbol{Z}^{(m+1)}) + \beta_{f0},\,
            \dots,\,
            \beta_{fL_f}(\boldsymbol{Z}^{(m+1)}) + \beta_{fL_f},
        \right)
    \end{aligned}
\end{equation}

where

\begin{equation}
    \begin{aligned}
        \alpha_{fl}(\boldsymbol{Z}^{(m+1)}) &= \sum_{i=1}^{n_1} \sum_{j=1}^{n_2} 
        \mathbb{I}(\gamma_{ij}^f = l) \cdot \mathbb{I}(Z_j^{(m+1)} = i), \\
        \beta_{fl}(\boldsymbol{Z}^{(m+1)}) &= \sum_{i=1}^{n_1} \sum_{j=1}^{n_2} 
        \mathbb{I}(\gamma_{ij}^f = l) \cdot \mathbb{I}(Z_j^{(m+1)} \neq i).
    \end{aligned}
\end{equation}

The Gibbs sampler can be computationally expensive, which can limit the Bayesian record linkage model to not very large datasets.  However, recent work \parencite[e.g., ][]{kundinger} relaxes the bipartite assumption to allow for much faster computation, allowing for Bayesian linkage to scale to millions of records.

\subsection{Post-Linkage Inference Framework of \textcite{slawski2024general}}
\label{sec:slawski}

\textcite{slawski2024general} developed a framework for inference about the parameters of statistical models estimated using linked data when mismatches are possible. The approach is specifically designed for secondary analysis settings where the analyst has a single linked file but not $D_1$ and $D_2$ themselves. 
For each linked pair in the linked dataset, we introduce an unobserved binary indicator to represent whether the pair constitutes a true match or a mismatch. Correctly linked pairs follow the substantive relationship of interest, e.g., a regression model relating $\boldsymbol Y$ to $\boldsymbol X$.  Mismatched pairs exhibit independence, e.g., $f(\boldsymbol Y | \boldsymbol X) = f(\boldsymbol Y)$, since these pairs correspond to unrelated individuals. This mixture model uses the relationship between $\boldsymbol Y$ and $\boldsymbol X$ to help distinguish genuine links and mismatches, which in turn 
reduces the influence of incorrect matches on parameter estimation.

Inference for parameters of interest is based on a composite likelihood \parencite{lindsay11988composite, varin2011overview} and an expectation-maximization (EM) algorithm.
In the expectation step (E-step), the algorithm computes the probability that each linked pair represents a true match, given the current parameter estimates and observed data. The maximization step (M-step) updates the model parameters by maximizing the expected complete-data likelihood, effectively weighting each observation by its estimated probability of being correctly matched. A key requirement of their approach is the availability of auxiliary match quality information for each pair (such as string similarity scores or probabilistic linkage weights), which is incorporated by modeling the match probabilities as functions of these auxiliary measures. 

The mixture model of \textcite{slawski2024general} presumes a single linked datafile and the presence of externally provided match quality measures.
We extend the approach to the multiple plausibly linked files that result from a Bayesian record linkage model.
We also provide methods that do not require externally provided measures of match quality.

\subsection{Multiple Imputation}
\label{sec:multi_impu}

Multiple imputation was developed initially as a strategy for handling missing values \parencite{rubin2018multiple}, although it has been applied to other contexts such as confidentiality protection \parencite{reiter:raghu}. In the context of missing data, we generate $M>1$ plausible completed datasets, $(D^{(1)}, \dots, D^{(M)})$.  We perform the statistical analysis of interest on each completed dataset, and combine the point and variances estimates via Rubin's rules \parencite{rubin:1987}.

Suppose we are interested in estimating some parameter $\theta$.  In each $D^{(m)}$ where $m=1, \dots, M$, let $\hat{\theta}^{(m)}$ and  $\hat{v}^{(m)}$ be the point estimate of $\theta$ and its associated variance estimate computed using $D^{(m)}$.
The multiple imputation point estimate of $\theta$ is 
\begin{equation}
    \bar{\theta} = \sum_{m=1}^{M} \hat{\theta}^{(m)}/M.\label{eq:MItheta}
\end{equation}
The multiple imputation variance estimate is  
\begin{equation}
    T = \bar{v} + (1+1/M)b,
\end{equation}
where 
\begin{align} \label{eq:MIv}
    \bar{v} &= \sum_{m=1}^M \hat{v}^{(m)}/M\\ \label{eq:MIb}
    b &= \sum_{m=1}^M (\hat{\theta}^{(m)}-\bar{\theta})^2/(M-1).
\end{align}
We make inferences using a $t$-distribution with $d$ degrees of freedom, $\bar{\theta} \pm t_{d}(\alpha/2)\sqrt{T}$, where 
\begin{equation}
    d = (M-1)\left(1+\frac{\bar{v}}{(1+1/M)b}\right)^2.\label{eq:MIdf}
\end{equation}

In our setting, $\boldsymbol{Z}$ is the missing data. Using the Bayesian record linkage model, we generate multiple imputations of $\boldsymbol Z$ in the Gibbs sampler.  We select $M$ of these draws, adequately spaced to ensure approximate independence of $\boldsymbol{Z}^{(1)}, \dots, \boldsymbol{Z}^{(M)}$. For each $\boldsymbol{Z}^{(m)}$, where $m=1,\dots,M$, we then create the linked datafile, $D^{(m)} = 
\left\{\left(\boldsymbol{x}_i, y_j\right): Z^{(m)}_j = i \leq n_1, j = 1, \dots, n_2\right\}$.  We also write $D^{(m)}$ as  $\{(\boldsymbol{x}_k^{(m)}, y_k^{(m)}): k = 1, \dots, n_{12}(\boldsymbol Z^{(m)})\}$, allowing $k$ to index the linked records in $\boldsymbol{Z}^{(m)}$.  For ease of notation, let $n^{(m)} = n_{12}(\boldsymbol Z^{(m)})$.

\section{Post-linkage Multiple Imputation Inference using Confidence Measures}\label{sec:PLMIc}
In this section, we present a method for multiple imputation inferences for the linear regression of $\boldsymbol Y$ on $\boldsymbol X$ using the $M$ plausible linked datasets generated from the Bayesian record linkage model in Section \ref{sec:BRLmodel}. We presume that the analyst has access to or can compute some measure of linkage quality; we remove this presumption in Section \ref{sec:PLMI}. 

Specifically, we leverage the strategy of  \textcite{slawski2024general} and propose a latent class model for the true linkage status of each linked record.  The latent class probabilities are conditional on the measure of linkage quality. We use an EM algorithm to estimate the parameters in the regression of $\boldsymbol Y$ on $\boldsymbol X$, along with estimated variances, and we combine the results using Rubin's rules from Section \ref{sec:multi_impu}. We call this model the post-linkage multiple imputation with confidence measures, which we abbreviate as PLMIc.

\subsection{Model Specification}\label{sec:PLMIc_model_specification}
 
For $m=1,\dots,M$ and $k=1, \dots n^{(m)}$, let $c_k^{(m)}$ be the analyst's confidence measure for record $k$ in $D^{(m)}$.  In our simulation studies, we set $c_k^{(m)}$ equal to the expression in \eqref{eq:confidence_measure} for the corresponding linked pair $(i,j)$, that is,  $c_k^{(m)} = c_{ij}^{(m)}$. This quantity is a multiple imputation version of the weights used in non-Bayesian approaches to probabilistic record linkage \parencite{fellegi1969theory}. 
We note that confidence measures can be defined in other ways.  For example, given enough $M$, the analyst could compute the percentage of times across $(D^{(1)}, \dots, D^{(m)})$ that each record pair is a match, and construct $c_k^{(m)}$ using those percentages.

For any record $k$ in $D^{(m)}$ constructed by linking record pairs $(i,j)$, i.e., ${Z}^{(m)}_{j}=i$, 
let $t_{k}$ be a latent indicator of whether pair $(i,j)$ is a true link; that is, $t_{k} = 1$ when pair $(i,j)$ is correctly linked and $t_{k}=0$ otherwise. 

To facilitate inference, we adopt three main assumptions. 
    \begin{enumerate}
        \item[A1.]  For any estimated linkage $\boldsymbol Z^{(m)}$, each $y_k$ is independent of $c^{(m)}_{k}$ given  $\boldsymbol{x}_{k}$ and $t_{k}$; that is, we have \\
 $                   f({y}_{k} \mid  \boldsymbol{x}_{k}, t_{k}, c^{(m)}_{k})
                    =
                    f({y}_{k} \mid \boldsymbol{x}_{k}, t_{k}).$
        \item[A2.]  For any estimated linkage $\boldsymbol Z^{(m)}$, ${y}_{k}$ follows the latent class model,
               \begin{equation}
                    {y}_{k} \mid \boldsymbol{x}_{k}, t_{k}
                    \sim 
                    \begin{cases}
            \phi({y}_{k} \mid \boldsymbol{x}_{k}; \boldsymbol{\theta}) & \text{if  } t_{k} = 1 \\
                        p_Y(y_{k}) & \text{if  } t_{k} = 0,
                    \end{cases}
                \end{equation}
             where $\phi({y}_{k} \mid \boldsymbol{x}_{k}; \boldsymbol{\theta})$ denotes the distribution of correctly linked responses conditional on the covariates with parameters  $\boldsymbol{\theta}$, and $p_Y$ is the marginal distribution of the responses.
        \item[A3.] For any estimated linkage $\boldsymbol Z^{(m)}$, the probability that record $k$ is a true link is independent of $\boldsymbol x_{k}$ given $c^{(m)}_{k}$, that is, 
               \begin{equation}
                    \Pr(t_{k}=1 \mid c^{(m)}_{k}, x_k) 
                    = \Pr(t_{k}=1 \mid c^{(m)}_{k})
                    = h(c^{(m)}_{k};\boldsymbol{\eta}),
               \end{equation} 
               for some monotonically increasing function $h:\, \mathbb{R} \to [0,1]$ with parameter $\boldsymbol{\eta}$.
    \end{enumerate}

Assumption A1 implies that if we know the true linking status $t_{k}$ of record $k$, the confidence measure used to  determine the estimated linkage status $Z^{(m)}_J$ no longer provides relevant information about the distribution of the response variable. This assumption seems plausible in general, since the confidence measures reflect the similarity of the linking variables in the record pair without any reliance on $\boldsymbol Y$ and $\boldsymbol X$.  On the other hand,  A1 could be violated if the confidence measure somehow encodes a covariate that is not included in $\boldsymbol X$ yet predicts $\boldsymbol Y$.  In this case, the analyst's model for $\boldsymbol Y | \boldsymbol X$ is mis-specified.  In such cases, the analyst may wish to add features to $\boldsymbol X$ that encode aspects of the linking variables.  This could require harmonization of variables used in $\boldsymbol X$ that appear in $D_1$ and $D_2$; for example, designate one of the files as having the true values of this common predictor.

Assumption A2 specifies the data generating process: if record $k$ is true link, then $y_k$ follows a regression on  $\boldsymbol{x}_k$ and, otherwise,  ${y}_k$ simply follows some marginal distribution.  Of course, the regression could be mis-specified, leading to a violation of A2 in the case of $t_{k}=1$.  In section \ref{sec:PLMI}, we discuss the benefits of identifying records where it is known that $Z_j=1$; these can inform the regression specification. For false links where $t_{k}=0$, A2 implies a different data generating process for responses. As we do not observe $\boldsymbol X$ for these cases (at least for covariates only in $D_1$), we are forced to presume the values follow some marginal distribution $p_Y$.  Under A1,  this distribution is independent of the confidence measures as well.

Assumption A3 implies that only the confidence measure is needed to estimate the true linkage status of each linked pair in $\boldsymbol Z^{(m)}$. That is, any two declared record pairs, say $(i,j)$ and $(i',j')$, with the same value of the  confidence measure have the same probability of being true links, regardless of their covariate values. This could be violated if, for some reason, the Bayesian record linkage is more likely to mismatch certain subsets of the study population defined by values of $\boldsymbol X$ than other subsets.  When A3 is violated, it suggests the record linkage model itself should be modified to allow different linkage probabilities for different covariates \parencite{wortmanthesis}.

In practice, $p_Y$ in A2 can be specified as any plausible marginal. It also could be estimated via some flexible density estimator \parencite{silverman2018density, sheather2004density, devroye1987course}.  Technically, this density should be the density of the non-linked records. For computational simplicity, we recommend estimating it as the marginal density of $\boldsymbol Y$ in $D_2$, before any record linkage is conducted. This simplification is reasonable when the response variables for the nonlinked records are not too dissimilar from the full set of records in $D_2$, e.g., the linking variables are uncorrelated with $\boldsymbol Y$.

A1, A2, and A3 do not preclude the analyst from using different model specifications in each $D^{(m)}$.  However, using different model specifications seems unlikely to be the case in practice, particularly since our primary goal is to learn about a specific  $\theta$. We presume the forms of $h$ are identical for all plausible datafiles.

\subsection{Inference Methods}\label{sec:PLMIc_inference}

We now present a strategy for multiple imputation inferences for the PLMIc model defined by A1--A3. We presume the Bayesian record linkage model from  Section~\ref{sec:BRLmodel} has been estimated, resulting in $M$ linked datafiles $D^{(1)}, \dots,  D^{(M)}$. Each $D^{(m)}$ includes a set of confidence measures, $\boldsymbol C^{(m)} = \{c^{(m)}_k: k = 1, \dots, n^{(m)}\}.$  The key step is to run an EM algorithm in each $D^{(m)}$ to obtain maximum likelihood estimates of all parameters, as well as their estimated variances, as we now describe.

For any $D^{(m)}$, we can write the log-likelihood of the model parameters $(\boldsymbol{\theta}, \boldsymbol{\eta})$ as
\begin{align}\label{eq:log_likelihood_CM}
    \ell(\boldsymbol{\theta}, \boldsymbol{\eta}; \mid D^{(m)})
    = \sum_{k=1}^{n^{(m)}} \log\left(\phi({y}_k^{(m)} \mid \boldsymbol{x}_k^{(m)}; \boldsymbol{\theta}) \, h(c_k^{(m)};\boldsymbol{\eta}) + p_{Y}({y}_k^{(m)}) \, (1 - h(c_k^{(m)};\boldsymbol{\eta}))\right).
\end{align}

Thus, letting $\boldsymbol{T}^{(m)} = (t_1^{(m)}, \dots, t_{n^{(m)}}^{(m)})$ the complete-data log-likelihood is
\begin{align}\label{eq:complete_data_log_likelihood_CM}
    \ell(&\boldsymbol{\theta}, \boldsymbol{\eta}, \boldsymbol{T}^{(m)}; D^{(m)}) 
    &= \sum_{k=1}^{n^{(m)}} t_k^{(m)}\log\big(\phi({y}_k^{(m)} \mid \boldsymbol{x}_k^{(m)}; \boldsymbol{\theta}) \,
        h(c_k^{(m)}; \boldsymbol{\eta})\big) + (1-t_k^{(m)})\log\big(p_Y({y}_k^{(m)}) \,
        (1 - h(c_k^{(m)}; \boldsymbol{\eta}))\big).
\end{align}

We now describe the expectation (E) step and maximization (M) step of the EM algorithm intended to maximize \eqref{eq:complete_data_log_likelihood_CM}.

\paragraph{E-Step.}
Let $\boldsymbol{\theta}_{\text{old}}^{(m)}$ and $\boldsymbol{\eta}_{\text{old}}^{(m)}$ be the current values of the parameters in the iterative algorithm.  For $k =1, \dots, n^{(m)}$, we have  $t_k^{(m)} \sim \text{Bernoulli}(\tilde{p}_k^{(m)})$,  with
\begin{align}\label{eq:t_hat_CM}
    \tilde{p}_k^{(m)} 
    &= \frac{
        \phi({y}_k^{(m)} \mid \boldsymbol{x}_k^{(m)}; \boldsymbol{\theta}_{\text{old}}^{(m)}) \,
        h(c_k^{(m)}; \boldsymbol{\eta}_{\text{old}}^{(m)})
    }{
        \phi({y}_k^{(m)} \mid \boldsymbol{x}_k^{(m)}; \boldsymbol{\theta}_{\text{old}}^{(m)}) \, h(c_k^{(m)}; \boldsymbol{\eta}_{\text{old}}^{(m)})
        + p_Y({y}_k^{(m)}) \, (1 - h(c_k^{(m)}; \boldsymbol{\eta}_{\text{old}}^{(m)})) 
    }.
\end{align}

Using \eqref{eq:complete_data_log_likelihood_CM} and \eqref{eq:t_hat_CM}, the expectation of the complete-data log-likelihood is
\begin{align}
    Q(&\boldsymbol{\theta}, \boldsymbol{\eta};\boldsymbol{\theta}_{\text{old}}^{(m)}, \boldsymbol{\eta}_{\text{old}}^{(m)}) \notag \\
    &= \sum_{k=1}^{n^{(m)}}
        \tilde{p}_k^{(m)}\big[\log h(c_k^{(m)}; \boldsymbol{\eta}) + \log \phi({y}_k^{(m)} \mid \boldsymbol{x}_k^{(m)}; \boldsymbol{\theta})\big]
        + (1-\tilde{p}_k^{(m)})\big[\log (1 - h(c_k^{(m)}; \boldsymbol{\eta}) + \log p_Y({y}_k^{(m)})\big].
\end{align}

\paragraph{M-Step.} 
We update  $\boldsymbol{\theta}_{\text{new}}^{(m)}$ and $\boldsymbol{\eta}_{\text{new}}^{(m)}$ by minimizing $Q(\boldsymbol{\theta}, \boldsymbol{\eta};\boldsymbol{\theta}_{\text{old}}^{(m)}, \boldsymbol{\eta}_{\text{old}}^{(m)})$ using the R function \texttt{optim}. Since $\boldsymbol{\theta}$ and $\boldsymbol{\eta}$ do not interact with each other in the $Q$ function, the optimization can be decomposed into two separate subproblems:
\begin{align}
\boldsymbol{\theta}_{\text{new}}^{(m)} &= \underset{\boldsymbol{\theta}}{\mathrm{argmin}}\,Q(\boldsymbol{\theta}, \boldsymbol{\eta}_{\text{old}}^{(m)};\boldsymbol{\theta}_{\text{old}}^{(m)}, \boldsymbol{\eta}_{\text{old}}^{(m)}), \\
\boldsymbol{\eta}_{\text{new}}^{(m)} &= \underset{\boldsymbol{\eta}}{\mathrm{argmin}}\,Q(\boldsymbol{\theta}_{\text{new}}^{(m)}, \boldsymbol{\eta};\boldsymbol{\theta}_{\text{old}}^{(m)}, \boldsymbol{\eta}_{\text{old}}^{(m)}).
\end{align}
Each subproblem is solved independently using \texttt{optim} function from \texttt{stats} package in R.

Let $\hat{\boldsymbol{\theta}}^{(m)}$ and $\hat{\boldsymbol{\eta}}^{(m)}$ be the values returned from the EM algorithm after it has been judged to converge. 
Under certain regularity conditions, the variance of the MLE is estimated with negative inverse Hessian of log-likelihood in \eqref{eq:log_likelihood_CM}, i.e.,
\begin{equation}\label{eq:variance_nonCM}
    \hat{\boldsymbol{V}}^{(m)} = -\left[D_{(\boldsymbol{\theta}, \boldsymbol{\eta})}^2 \, \ell(\boldsymbol{\theta}, \boldsymbol{\eta}; \boldsymbol{Y}^{(m)} \mid \boldsymbol{X}^{(m)}, \boldsymbol{C}^{(m)}, \boldsymbol{Z}^{(m)})\right]^{-1} \Big|_{(\boldsymbol{\theta}, \boldsymbol{\eta}) = (\hat{\boldsymbol{\theta}}^{(m)}, \hat{\boldsymbol{\eta}}^{(m)})}.
\end{equation}
In practice, the Hessian matrix is computed numerically using the \texttt{hessian} function from the \texttt{numDeriv} package in R, evaluated at the converged parameter estimates $(\hat{\boldsymbol{\theta}}^{(m)}, \hat{\boldsymbol{\eta}}^{(m)})$.

For any element of the vector $\hat{\boldsymbol \theta}^{(m)}$, say $\hat{\theta}^{(m)}_l$, let  $\hat v_l^{(m)}$ be the $l$-th diagonal element of $\hat{\boldsymbol{V}}^{(m)}$.
The multiple imputation inferences for $\theta_l$  then use 
$\big\{\hat{\theta}_l^{(m)}, \hat v_l^{(m)}: m=1, \dots, {M}\big\}$ in \eqref{eq:MItheta} -- \eqref{eq:MIdf}.

\section{Post-linkage Multiple Imputation Inference Without Confidence Measures}\label{sec:PLMI}

We now present a multiple imputation approach that, unlike PLMIc, is used without confidence measures. We refer to this model as the post-linkage multiple imputation model, abbreviated PLMI.  In lieu of confidence measures, the PLMI model requires that some record pairs are known with certainty (or at least treated as such) to be true links. We call these records seeds \parencite{dalzell:reiter}.

\subsection{Model Specification}\label{sec:PLMI_model_specification}

We presume that  $n_s < n_2$ records in $D_2$ are seeds.  Therefore, when fitting the Bayesian record linkage model of Section \ref{sec:BRLmodel}, for all seeds $j$  we preset $Z_j$ to equal the index of its link in $D_1$.  Thus, each $D^{(m)}$ generated from estimating the Bayesian record linkage comprises $n_s$ seeds and $n^{(m)} - n_s$ pairs with unknown linkage status.  For the seeds, we set $t^{(m)}_k=1$ for all $m$.  For the uncertain links, we treat $t^{(m)}_k$ as unknown. 

To specify the PLMI model, we assume A2 from Section \ref{sec:PLMIc} and modify A3 to recognize the method does not use confidence measures.  This modification is Assumption A4.

\begin{enumerate}
 \item[A4.] For any estimated linkage $\boldsymbol Z^{(m)}$, the probability that record $k$ is a true link is independent of $\boldsymbol x_{k}$, that is, 
       
               \begin{equation}
                    \Pr(t_{k}=1 \mid \boldsymbol x_k) 
                    = \Pr(t_{k}=1)
                    = \delta.
               \end{equation} 
\end{enumerate}     
Assumption A4 can be interpreted similarly to A3, except it requires an equal probability that each link in $D^{(m)}$ is correct for purposes of estimating the latent $t_k$.  This is a stronger assumption than A3.  Indeed, absent the use of confidence measures, the data may provide only weak information for identifying which links in any $\boldsymbol Z^{(m)}$ are likely correct and which are likely incorrect.  This motivates the need for seeds, which help pin down the regression model estimates and therefore information to identify the likely false links.

\subsection{Inference Methods}\label{sec:PLMI_inference}
For multiple imputation inferences about $\boldsymbol \theta$, we follow a similar strategy to what we used in Section \ref{sec:PLMIc_inference} for the PLMIc model. The main differences are that the EM algorithm no longer includes $\boldsymbol C^{(m)}$ as conditioning information for $t^{(m)}_k$ and that $t^{(m)}_k=1$ for all seeds.

To write the log-likelihood for the model implied by A2 and A4, for convenience we presume each $D^{(m)}$ has been ordered so that the first $n_s$ records are seeds.  Thus, the log-likelihood for $(\boldsymbol{\theta}, \delta)$ is 
\begin{align}\label{eq:log_likelihood_NC}
    \ell(\boldsymbol{\theta}, \delta; D^{(m)})
    = n_s \log \delta 
    + \sum_{k=1}^{n_s} \log \phi({y}_k \mid \boldsymbol{x}_k;\, \boldsymbol{\theta}) + \sum_{k=n_s+1}^{n^{(m)}} \log \left( 
        \phi({y}_k^{(m)} \mid \boldsymbol{x}_k^{(m)};\, \boldsymbol{\theta})\, \delta 
        + p_Y({y}_k^{(m)})\, (1 - \delta) 
    \right).
\end{align}
Thus, the complete-data log-likelihood is
\begin{align}\label{eq:complete_data_log_likelihood}
    \ell\Big(&\boldsymbol{\theta}, \delta, \boldsymbol T^{(m)};\, D^{(m)}\Big) 
    =  n_s \log \delta + \sum_{k=1}^{n_s} \log \phi({y}_k \mid \boldsymbol{x}_k;\, \boldsymbol{\theta}) \notag \\ 
    &+ \sum_{k=n_s+1}^{n^{(m)}} 
     t_k^{(m)} \log \left( \phi({y}_k^{(m)} \mid \boldsymbol{x}_k^{(m)}; \boldsymbol{\theta}) \delta \right)
  + (1-t_k^{(m)})\log\left( p_Y({y}_k^{(m)}) (1 - \delta) \right).
\end{align}

We obtain the maximum likelihood estimate of $\boldsymbol \theta$ by iteratively performing the expectation (E) step and maximization (M) step as follows.  To save space, we present the E step for PLMI here.

\paragraph{E-Step} Given current values  $\boldsymbol{\theta}_{\text{old}}^{(m)}$ and $\delta_{\text{old}}^{(m)}$, we have $t_k^{(m)}\sim \text{Bernoulli}(\tilde{p}_k^{(m)})$, where $n_s +1 \leq k \leq n^{(m)}$ where 
\begin{align}\label{eq:t_hat}
    \tilde{p}_k^{(m)}
    = \frac{
        \phi({y}_k^{(m)} \mid t_k^{(m)}=1, \boldsymbol{x}_k^{(m)}; \boldsymbol{\theta}_{\text{old}}^{(m)}) \, \delta_{\text{old}}^{(m)}
    }{
        \phi({y}_k^{(m)} \mid t_k^{(m)}=1, \boldsymbol{x}_k^{(m)}; \boldsymbol{\theta}_{\text{old}}^{(m)}) \, \delta_{\text{old}}^{(m)} + p_Y({y}_k^{(m)})\,(1-\delta_{\text{old}}^{(m)})
    }.
\end{align}

Using  \eqref{eq:complete_data_log_likelihood} and \eqref{eq:t_hat}, the expectation of complete-data likelihood is
\begin{align}
Q(\boldsymbol{\theta}, \delta;\boldsymbol{\theta}_{\text{old}}^{(m)}, \delta_{\text{old}}^{(m)})
&= n_s \log \delta 
  + \sum_{k=1}^{n_s} \log \phi({y}_k \mid \boldsymbol{x}_k;\, \boldsymbol{\theta}) \notag \\
&\quad + \sum_{k=1}^{n^{(m)}} 
      \Big( \tilde{p}_k^{(m)} \log\big(\delta \,\phi({y}_k^{(m)} \mid \boldsymbol{x}_k^{(m)};\, \boldsymbol{\theta})\big) 
      + (1-\tilde{p}_k^{(m)}) \log\big((1-\delta)\,p_Y({y}_k^{(m)})\big) \Big).
\end{align}

We use an M step that is similar the one used in Section \ref{sec:PLMIc_inference}, but based on \eqref{eq:complete_data_log_likelihood} of course.  After convergence of the EM algorithm, we construct the covariance matrix $\hat{\boldsymbol V}^{(m)}$ using the Hessian as in \eqref{eq:variance_nonCM}.  Finally, we collect the values of $(\hat \theta^{(1)}_1, \hat v^{(m)}_1), \dots, (\hat \theta^{(M)}_l, \hat v^{(M)}_l)$, which we use for multiple imputation inferences.

\section{Simulation Studies}\label{sec:simu_study}
In this section, we evaluate the performances of PLMIc and PLMI models using simulation studies. In section~\ref{sec:simu_setting}, we describe the simulation design. In section ~\ref{sec:simu_PLMIc} and section ~\ref{sec:simu_PLMI}, we present results of the simulations for PLMIc and PLMI models,  respectively. In section~\ref{sec:simu_example}, we dive into a specific simulation run to illustrate how our algorithm works in more detail. In section~\ref{sec:simu_violation_indep} we test the robustness of the models' performances when there is dependence among the covariates and the linkage statuses. For ease of explanation and computation, the simulations in these sections all use a single, normally distributed covariate.  In section~\ref{sec:simu_multiReg}, we illustrate how to extend the models to multiple regression and use a non-normal marginal distribution of the response variable.

\subsection{Simulation Settings}\label{sec:simu_setting}

We base our main set of simulations on the synthetic datasets developed by \textcite{sadinle2017bayesian} to evaluate the effectiveness of the Bayesian record linkage model alone. In these datasets, $D_1$ and $D_2$ each comprise $n_1=n_2=500$ records. There are no duplicates within any file, but there are matching records across the files. Specifically, $n_{12}$ records in $D_2$ have true links in $D_1$ and $n_2 - n_{12}$ records in $D_2$ do not have true links in $D_1$. The value of $n_{12}$ is either 250 or 500 representing two settings, one with 50\% overlap and one with 100\% overlap. Each file includes $F=4$ linking variables, namely a person's first name, last name, age, and occupation. In addition, $D_1$ contains a predictor variable $x$ and $D_2$ contains a response variable $y$.

As described by \textcite{sadinle2017bayesian}, 
the first and last names are sampled from frequency tables compiled from Australian public data sources (e.g., telephone directories), so that common names appear with higher probability. This sampling scheme reflects real-world distributions of name frequencies. Age and occupation are generated in a similar fashion, using marginal or joint distributions derived from genuine data. These true values are then distorted by injecting errors into a subset of the fields. Each record has a fixed number of erroneous fields $n_{\text{error}}$ selected uniformly at random.  The types of errors include typographical errors, phonetic substitutions, optical character recognition (OCR) errors, etc. These distortions are intended to simulate realistic noise commonly found in administrative data. For further details on the data generation process, see \textcite{christen2009accurate} and \textcite{christen2013flexible}.  We consider two scenarios: one with a low error rate where only one field contains errors, denoted as $n_{\text{error}}=L$, and one with a high error rate where three fields contain errors, denoted as $n_{\text{error}}=H$.

For age and occupation, which we index with $f=3$ and $f=4$, we use indicators of exact match. When the age for record $j$ in $D_2$ is equal to the age for record $i$ in $D_1$, we set $\gamma^3_{ij} = 1$; otherwise, we set  $\gamma^3_{ij} = 0$.  Likewise for occupation, we set $\gamma^4_{ij} = 1$ when the two occupations match and  $\gamma^4_{ij} = 0$ otherwise. For first and last names, which we index with $f=1$ and $f=2$, we compute the standardized Levenshtein distances \parencite{winkler1990string} of the names for each record pair $(i,j)$. From these we construct a similarity score, which ranges from $[0,1]$ whereby lower scores correspond to smaller distances (greater similarity) between the names.  We construct $\gamma^1_{ij}$ and $\gamma^2_{ij}$ based on binned versions the similarity scores, using bins $\{1,2,3,4\}$ corresponding in order to exact matches with score equal to $0$ and inexact matches with scores in $(0,0.25], (0.25,0.5]$, and $(0.5, 1]$. 

We are interested in a linear regression model, $y = \beta_0 + \beta_1 x + \varepsilon$, with $\varepsilon \sim \mathcal{N}(0, \sigma^2)$. We sample each $x$ from the standard normal distribution. We set $\beta_0=\beta_1=3$. We set $\sigma^2$ to yield coefficients of determination $R^2 \in \{0.3, 0.6, 0.9\}$. The regression coefficients are held constant across all scenarios to ensure comparability, as false links pull the estimates of $\beta_1$ toward zero, reducing the bias of estimates inherently. Each scenario is simulated 500 times.

In addition to PLMIc and PLMI, in each simulation run we estimate two other models for comparisons. First, we estimate an ordinary least squares (OLS) model using only true links in $D_1$ and $D_2$. We refer to this as Perfect, as this model offers the best estimates of the coefficients we could have given the overlap (and if we ignore the marginal information of unpaired records). Second, we estimate a two-step model, which we refer to as TS.OLS. TS.OLS estimates an OLS with linked pairs in each $D^{(m)}$, and combines the estimates using the multiple imputation rules. Hence, TS.OLS does not use the information from the regression relationship to help decide if certain pairs in $D^{(m)}$ are likely true or false links.  Finally, for TS.OLS, PLMI, and PLMIc, we use the same draws from the Gibbs sampler for the Bayesian record linkage model, which we estimate with $1000$ runs with the first $100$ as burn-in.  Diagnostic checks offer no evidence that the samplers fail to converge.

\subsection{Results for the PLMIc Model}\label{sec:simu_PLMIc}

To estimate PLMIc, we specify $h$ in Assumption A3 in each $D^{(m)}$, where $m =1, \dots, M$, using a logistic regression of $\boldsymbol T^{(m)}$ on $\boldsymbol C^{(m)}$, so that  
\begin{equation}
    h(c_k^{(m)};\boldsymbol{\eta}) = \frac{1}{1+\exp\left(-(\eta_0 +\eta_1 c_k^{(m)})\right)}.\label{eq:hinPMLIc}
\end{equation}
For computational simplicity, we estimate $\boldsymbol{\eta}$ separately in each $D^{(m)}$.  The specification in \eqref{eq:hinPMLIc} 
satisfies the requirement in A3 of being monotonically increasing with range $[0,1]$. The logistic function ensures that the latent class probability  approaches one smoothly as $c_k^{(m)}$ becomes large. This aligns with the intuition that incremental evidence, e.g., additional matching fields when $F$ is already large, may not increase the likelihood that a record pair is a true link when we already have high confidence in that possibility.

For $p_Y$, we use a normal distribution as $\boldsymbol Y$ is marginally normal,  We estimate the mean and variance of this marginal distribution  using all response values in $D_2$. 

We initialize the EM algorithm using the TS.OLS estimates at each iteration.  We also consider situations where we want to run PLMIc and additionally have seeds, in which case we initialize the EM algorithm at the OLS estimates obtained from regression $\boldsymbol Y$  on $\boldsymbol X$ using only the seeds.  The latter strategy has been shown in other contexts to improve the convergence of EM algorithms \parencite{yi2014alternating, yi2015regularized, yi2016solving, balakrishnan2017statistical, klusowski2019estimating}. We use all $900$ iterations of the Bayesian record linkage after the burn-in period to generate $D^{(1)}, \dots, D^{(M)}$; hence, we set $M=900$ in these simulations. Although the resulting $D^{(m)}$ may not be independent, this dependence does not invalidate the multiple imputation inferences when $M$ is sufficiently large \parencite{hu2013independent}.

\begin{table}[t]
\centering
\caption{Simulation results when running PLMIc, TS.OLS, and Perfect. The column $n_{\text{error}}$ refers to high and low numbers of errors in the linking variables; $R^2$ refers to the coefficient of determination in the regression; and, the overlap refers to the percentage of records that are true links.  Results based on 500 simulation runs.}
\label{tab:SimuRes_ConfidenceMeasure}

\resizebox{\textwidth}{!}{
\begin{tabular}{@{}rrrrrrrrrrrr@{}}
\toprule
\textbf{Parameter} & $n_{\text{error}}$ & $R^2$ & \multicolumn{3}{c}{\textbf{Coverage (\%)}} & \multicolumn{3}{c}{\textbf{Median Absolute Diff. ($\times 100$)}} & \multicolumn{3}{c}{\textbf{Median Length ($\times 100$)}} \\ \cmidrule(lr){4-6} \cmidrule(lr){7-9} \cmidrule(lr){10-12}
                   &                    &       & PLMIc & TS.OLS & Perfect & PLMIc & TS.OLS & Perfect & PLMIc & TS.OLS & Perfect \\ \midrule
\multicolumn{12}{l}{\textbf{Panel A:} overlap = 50\%} \\ \midrule
\multirow{6}{*}{$\beta_1$} 
& \multirow{3}{*}{L} & 0.3 & 86.6 & 89.6 & 89.2 & 43.7 & 22.8 & 22.3 & 203.4 & 99.4 & 96.0 \\
&                     & 0.6 & 93.8 & 86.4 & 89.6 & 12.8 & 13.8 & 10.6 & 62.6  & 57.4 & 51.2 \\
&                     & 0.9 & 91.0 & 76.8 & 89.4 & 4.5  & 10.9 & 4.4  & 21.7  & 33.9 & 20.9 \\ \cmidrule(lr){2-12}
& \multirow{3}{*}{H} & 0.3 & 89.6 & 84.2 & 91.4 & 36.2 & 27.1 & 18.1 & 170.3 & 108.2 & 96.3 \\
&                     & 0.6 & 94.0 & 67.2 & 89.8 & 11.3 & 27.7 & 9.7  & 63.7  & 68.3 & 51.2 \\
&                     & 0.9 & 92.6 & 40.0 & 92.6 & 4.4  & 25.7 & 4.2  & 23.0  & 47.7 & 20.9 \\ \midrule
\multirow{6}{*}{$\beta_0$} 
& \multirow{3}{*}{L} & 0.3 & 92.6 & 93.0 & 91.8 & 33.1 & 16.9 & 17.7 & 205.9 & 97.2 & 95.5 \\
&                     & 0.6 & 94.4 & 92.2 & 90.4 & 11.4 & 9.3  & 9.4  & 60.4  & 54.7 & 51.0 \\
&                     & 0.9 & 92.4 & 97.8 & 91.4 & 4.1  & 4.3  & 4.0  & 21.6  & 29.4 & 20.8 \\ \cmidrule(lr){2-12}
& \multirow{3}{*}{H} & 0.3 & 96.8 & 93.4 & 89.2 & 24.8 & 20.9 & 19.8 & 162.6 & 102.8 & 96.1 \\
&                     & 0.6 & 94.2 & 94.6 & 90.0 & 11.3 & 10.3 & 10.1 & 62.0  & 60.9 & 51.1 \\
&                     & 0.9 & 92.6 & 99.2 & 88.4 & 4.8  & 5.3  & 4.4  & 23.2  & 37.7 & 20.8 \\ \midrule
\multicolumn{12}{l}{\textbf{Panel B:} overlap = 100\%} \\ \midrule
\multirow{6}{*}{$\beta_1$} 
& \multirow{3}{*}{L} & 0.3 & 90.4 & 90.4 & 90.6 & 15.3 & 14.5 & 14.9 & 67.4 & 67.6 & 67.5 \\
&                     & 0.6 & 91.4 & 90.6 & 91.0 & 7.8  & 8.2  & 7.8  & 36.0 & 36.0 & 36.0 \\
&                     & 0.9 & 90.4 & 86.0 & 89.6 & 3.1  & 3.4  & 3.1  & 14.7 & 15.1 & 14.7 \\ \cmidrule(lr){2-12}
& \multirow{3}{*}{H} & 0.3 & 89.2 & 88.0 & 88.6 & 13.9 & 14.0 & 14.2 & 67.7 & 68.0 & 67.6 \\
&                     & 0.6 & 90.0 & 87.4 & 90.2 & 7.5  & 8.3  & 7.7  & 36.3 & 36.7 & 36.2 \\
&                     & 0.9 & 90.4 & 82.0 & 90.8 & 3.2  & 3.9  & 3.1  & 14.8 & 16.0 & 14.8 \\ \midrule
\multirow{6}{*}{$\beta_0$} 
& \multirow{3}{*}{L} & 0.3 & 91.6 & 92.4 & 92.6 & 12.2 & 12.2 & 12.1 & 67.3 & 67.4 & 67.4 \\
&                     & 0.6 & 89.0 & 88.8 & 88.6 & 7.4  & 7.2  & 7.3  & 35.9 & 36.3 & 36.0 \\
&                     & 0.9 & 90.8 & 92.0 & 91.2 & 3.0  & 3.0  & 2.9  & 14.7 & 15.0 & 14.7 \\ \cmidrule(lr){2-12}
& \multirow{3}{*}{H} & 0.3 & 90.4 & 91.0 & 91.0 & 13.0 & 13.5 & 13.3 & 67.4 & 67.6 & 67.5 \\
&                     & 0.6 & 92.2 & 93.4 & 93.0 & 6.7  & 6.8  & 6.8  & 36.2 & 36.6 & 36.2 \\
&                     & 0.9 & 90.2 & 92.4 & 90.8 & 3.0  & 3.0  & 3.0  & 14.8 & 15.8 & 14.8 \\ \bottomrule
\end{tabular}
}
\end{table}

Table \ref{tab:SimuRes_ConfidenceMeasure} displays the results of the simulation for the different scenarios and models. The evaluation metrics include the median absolute difference between point estimators and ground truth, the empirical coverage rates, and median lengths of the 90\% confidence intervals. We report the median rather than the mean of the absolute differences because the EM algorithm can be sensitive to initialization and may occasionally converge to local optima, resulting in outlier estimates. In practice, this issue can be mitigated by running the algorithm multiple times with different initial values. However, since we did not implement such a strategy in our simulation study, we use the median as a robust summary statistic that is less sensitive to these occasional poor convergence outcomes. PLMIc exhibits coverage rates near the nominal $90\%$ level across all scenarios.
In contrast, TS.OLS exhibits coverage rates that deteriorate as the number of erroneous fields and the regression strength increase. 
For example, with high error rates, $R^2=0.9$, and 50\% overlap, 
the coverage rate for $\beta_1$ under TS.OLS is only 40\%. TS.OLS does not adequately account for the biases in point estimates that can arise when estimating the regression with many incorrect links. In contrast, PLMIc inflates the intervals, leading to better coverage rates.

Turning to the point estimates, we see that PLMIc tends to have nearly the same or lower absolute biases than TS.OLS in most scenarios, often approaching the absolute differences for the Perfect linkage.  when $R^2$ equals 0.6 or 0.9 in the 50\% overlap scenarion, PLMIc substantially reduces the absolute errors in estimates of $\beta_1$ compared to TS.OLS.  On the other hand, when the regression model has $R^2=0.3$ in these situations, PLMIc has larger absolute biases than TS.OLS.  This appears to result mainly from increased variance, as apparent in the increased interval lengths for PLMIc for this case.   

Overall, Table \ref{tab:Simu_Indep_Violation} suggests that PLMIc is best able to differentiate true and false matches when the regression is stronger, i.e., high $R^2$, and estimated with more cases, i.e., higher overlap. When the signal is weak with relatively few cases to estimate the regression, e.g., $R^2=0.3$ in the 50\% overlap and high error scenario, the EM algorithm sometimes finds local rather than global maxima, as the likelihood function is relatively flat in this scenario. Analysts can mitigate this problem to some extent by running the EM algorithm with different initializations and for longer times.

\begin{table}[t]
\centering
\caption{Simulation results for PLMIc and TS.OLS when we add seeds comprising 1\%, 5\%, or 10\% of the records in $D_2$. 
Results based on two scenarios with 50\% overlap, one with low error and $R^2 = 0.3$ and one with high error and $R^2 = 0.9$. Results based on 500 simulation runs.}
\label{tab:SimuRes_SeedProportion}
\small
\newcolumntype{R}{>{\raggedleft\arraybackslash}X}
\begin{tabular}{ccrrrrrrrrr}
\toprule
\textbf{Parameter} & \textbf{Seed Prop.} & \multicolumn{3}{c}{\textbf{Coverage (\%)}} & \multicolumn{3}{c}{\textbf{Median Absolute Diff. ($\times 100$)}} & \multicolumn{3}{c}{\textbf{Median Length ($\times 100$)}} \\ \cmidrule(lr){3-5} \cmidrule(lr){6-8} \cmidrule(lr){9-11}
& & PLMIc & TS.OLS & Perfect & PLMIc & TS.OLS & Perfect & PLMIc & TS.OLS & Perfect \\ \midrule
\multicolumn{11}{l}{\textbf{Panel A: $n_{\text{error}}=L$, $R^2=0.3$}} \\ \midrule
\multirow{3}{*}{$\beta_1$} & 1\% & 87.6 & 91.0 & 92.6 & 21.1 & 20.0 & 20.0 & 98.8 & 98.6 & 95.1 \\
& 5\% & 88.2 & 88.8 & 90.0 & 20.6 & 20.5 & 18.5 & 98.7 & 99.6 & 96.7 \\
& 10\% & 88.8 & 89.2 & 89.6 & 21.2 & 21.7 & 19.4 & 97.8 & 98.6 & 95.6 \\ \midrule
\multirow{3}{*}{$\beta_0$} & 1\% & 89.4 & 90.8 & 90.0 & 18.2 & 18.2 & 19.4 & 96.8 & 97.0 & 95.4 \\
& 5\% & 89.2 & 89.6 & 88.2 & 19.6 & 19.4 & 20.0 & 97.0 & 97.6 & 95.9 \\
& 10\% & 91.4 & 92.2 & 90.0 & 17.9 & 18.0 & 18.7 & 96.8 & 97.4 & 95.6 \\ \midrule
\multicolumn{11}{l}{\textbf{Panel B: $n_{\text{error}}=H$, $R^2=0.9$}} \\ \midrule
\multirow{3}{*}{$\beta_1$} & 1\% & 93.4 & 36.8 & 92.6 & 4.5 & 26.6 & 4.4 & 22.2 & 48.4 & 20.7 \\
& 5\% & 90.6 & 34.8 & 90.0 & 4.4 & 28.6 & 4.0 & 22.6 & 49.5 & 21.1 \\
& 10\% & 90.2 & 27.6 & 89.7 & 4.4 & 30.0 & 4.2 & 22.4 & 50.0 & 20.8 \\
\midrule
\multirow{3}{*}{$\beta_0$} & 1\% & 91.2 & 98.0 & 90.0 & 4.1 & 5.3 & 4.2 & 22.5 & 38.2 & 20.8 \\
& 5\% & 89.4 & 99.0 & 88.1 & 4.5 & 5.6 & 4.3 & 22.6 & 38.8 & 20.9 \\
& 10\% & 92.2 & 99.4 & 90.0 & 4.0 & 4.8 & 4.0 & 22.7 & 39.4 & 20.8 \\
\bottomrule
\end{tabular}
\end{table}

One way to potentially improve the performance of PLMIc is to utilize seeds. Seeds have the effect of anchoring the parameter estimates from the EM algorithm parameters near reasonable values. This can accelerate convergence and reduce the risk of finding local maxima. To examine this potential benefit, we conduct simulations in two scenarios where PLMIc performs noticeably better ($n_{\text{error}}=L$, $R^2 = 0.3$, $overlap = 0.5$) and worse ($n_{\text{error}}=H$, $R^2 = 0.9$, $overlap = 0.5$) than TS.OLS, varying seed proportions to 1\%, 5\%, and 10\% of the total records in $D_2$.  We utilize the seeds when estimating both PLMIc and  TS.OLS.

Table \ref{tab:SimuRes_SeedProportion} summarizes the results of 500 runs of the simulations of PLMIc with seeds. We see significant benefits of incorporating seed information. As evident in low $R^2$ and low error simulation results from Table \ref{tab:SimuRes_SeedProportion}, the inclusion of just 1\% seed information--- equivalent to only 5 known matches among the 500 records---yields a substantial improvement for PLMIc.  Its results are now essentially equivalent to those from TS.OLS in this simulation scenario. We also see that the PLMIc continues to outperform TS.OLS in the setting of $R^2=0.9$ with high error. In fact, the seeds seem not to help TS.OLS particularly much.

\subsection{Results for the PLMI Model}\label{sec:simu_PLMI}

PLMI requires seeds to guide the EM algorithm to reasonable estimates. In these simulations, we designate $5\%$ of the records in $D_2$ as seeds, i.e., $25$ known links out of $500$ records.  For brevity, we consider only 50\% overlap, as these scenarios are more challenging for the linkage methods.  

Table~\ref{tab:model_without_confmeas} displays the results of 500 simulation runs. PLMI attains reasonable coverage rates across all scenarios, with lowest rate of 82\% in the case of high error with $R^2=0.3$. TS.OLS again exhibits substantial undercoverage that worsens as the relationship between $\boldsymbol{x}$ and $\boldsymbol{y}$ strengthens and the number of erroneous linkage fields increases. For instance, under the scenario with high errors and $R^2 = 0.9$, the coverage rate for $\beta_1$ for TS.OLS is only 34\%. PLMI offers median absolute errors and average interval lengths that are quite similar to those for the Perfect model.  Evidently, the  seeds provide enough information for the model to estimate the regression about as effectively as using the true links.

\begin{table}[t]
\centering
\caption{Simulation results for PMLI, TS.OLS, and Perfect with $5\%$ seeds and 50\% overlap. Results based on 500 simulation runs.}
\label{tab:model_without_confmeas}
\small
\newcolumntype{R}{>{\raggedleft\arraybackslash}X}
\begin{tabularx}{\textwidth}{@{}R R R *{9}{R} @{}}
\toprule
\textbf{Parameter} & $n_{\text{error}}$ & $R^2$ (\%) 
& \multicolumn{3}{c}{\textbf{Coverage} (\%)} 
& \multicolumn{3}{c}{\textbf{Median Absolute Diff.} ($\times 100$)} 
& \multicolumn{3}{c}{\textbf{Median Length} ($\times 100$)} \\ \cmidrule(lr){4-6} \cmidrule(lr){7-9} \cmidrule(lr){10-12}
& & & PLMI & TS.OLS & Perfect & PLMI & TS.OLS & Perfect & PLMI & TS.OLS & Perfect \\ \midrule
\multirow{6}{*}{$\beta_1$} 
& \multirow{3}{*}{L} & 0.3 & 92.0 & 93.0 & 91.0 & 21.2 & 21.0 & 20.3 & 97.1 & 99.0 & 96.0 \\
& & 0.6 & 90.0 & 89.0 & 90.0 & 10.8 & 14.4 & 9.9 & 53.4 & 57.9 & 51.4 \\
& & 0.9 & 86.0 & 71.0 & 90.0 & 4.0 & 11.1 & 4.0 & 21.0 & 33.5 & 20.9 \\ \cmidrule{2-12}
& \multirow{3}{*}{H} & 0.3 & 82.0 & 82.0 & 88.0 & 26.8 & 29.9 & 18.5 & 1.06 & 1.08 & 96.7 \\
& & 0.6 & 87.0 & 68.0 & 87.0 & 12.0 & 27.4 & 10.8 & 0.59 & 68.3 & 51.2 \\
& & 0.9 & 88.0 & 34.0 & 86.0 & 4.1 & 28.7 & 4.1 & 22.5 & 49.6 & 20.8 \\ \midrule
\multirow{6}{*}{$\beta_0$} 
& \multirow{3}{*}{L} & 0.3 & 94.0 & 94.0 & 91.0 & 19.3 & 19.2 & 21.0 & 95.3 & 97.2 & 95.7 \\
& & 0.6 & 94.0 & 94.0 & 88.0 & 9.6 & 9.3 & 9.8 & 51.8 & 55.1 & 51.3 \\
& & 0.9 & 94.0 & 97.0 & 95.0 & 4.4 & 4.7 & 4.2 & 21.0 & 29.5 & 20.8 \\ \cmidrule{2-12}
& \multirow{3}{*}{H} & 0.3 & 91.0 & 91.0 & 88.0 & 19.8 & 19.6 & 20.0 & 100.6 & 102.6 & 95.9 \\
& & 0.6 & 90.0 & 93.0 & 87.0 & 10.2 & 10.3 & 10.7 & 55.8 & 61.1 & 51.3 \\
& & 0.9 & 91.0 & 100.0 & 91.0 & 4.4 & 5.0 & 4.1 & 22.6 & 38.9 & 20.7 \\ \bottomrule
\end{tabularx}
\end{table}

\subsection{Visualizing the Effects of PLMIc and PLMI on Linkage Quality}\label{sec:simu_example}

The EM algorithms in PLMI and PLMIc iteratively estimate the latent class probabilities for each potential record in each $D^{(m)}$ based on the consistency of the record's $(\boldsymbol x^{(m)}_k, y^{(m)}_k)$ with the regression. Links that do not accord with the estimated regression are assigned relatively low probabilities of being a true link, and links that align with the estimated regression get relatively higher probabilities. To illustrate this behavior, we examine the latent class probabilities for PLMIc and PLMI in a representative simulation run of the scenario with high error, $R^2=0.9$ and 50\% overlap.

\begin{figure}[t]
    \centering
    \includegraphics[width=0.9\linewidth]{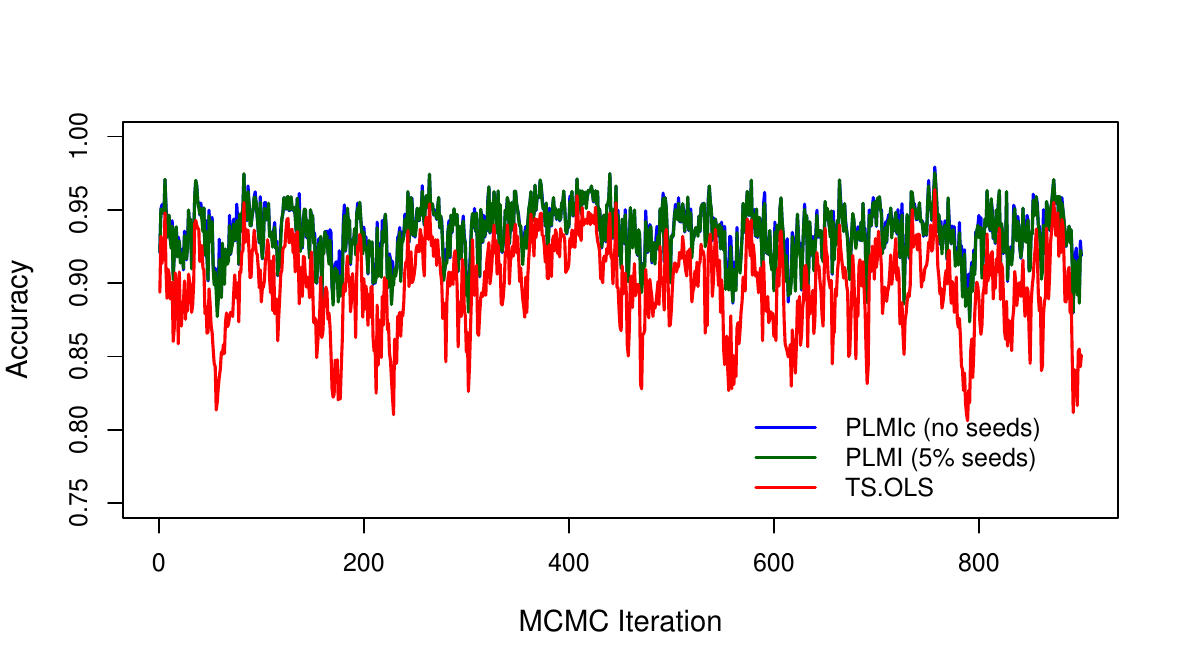}
    \caption{Percentage of correct links among the total number of links for PLMIc, PLMI, and TS.OLS.  Results are for simulation run and display the accuracy rates across iterations of the Gibbs samplers.}
\label{fig:accuracy_single_simulation}
\end{figure}

Figure \ref{fig:accuracy_single_simulation} displays the ratio of correct links over the total number of links for PLMIc, PLMI, and TS.OLS. For TS.OLS, the total number of links in any iteration is based on the set of links from the Bayesian record linkage model. 
For PLMIc and PLMI, the total number of links in any iteration is based on the set of records with latent class probabilities exceeding 0.5. The patterns in Figure \ref{fig:accuracy_single_simulation} are insensitive to modest perturbation of this threshold. 
Overall, both PLMIc and PLMI provide a higher fraction of correct links than TS.OLS, which helps explain its improved performance in this setting.

\begin{figure}[h]
    \centering
\includegraphics[width=0.9\linewidth]{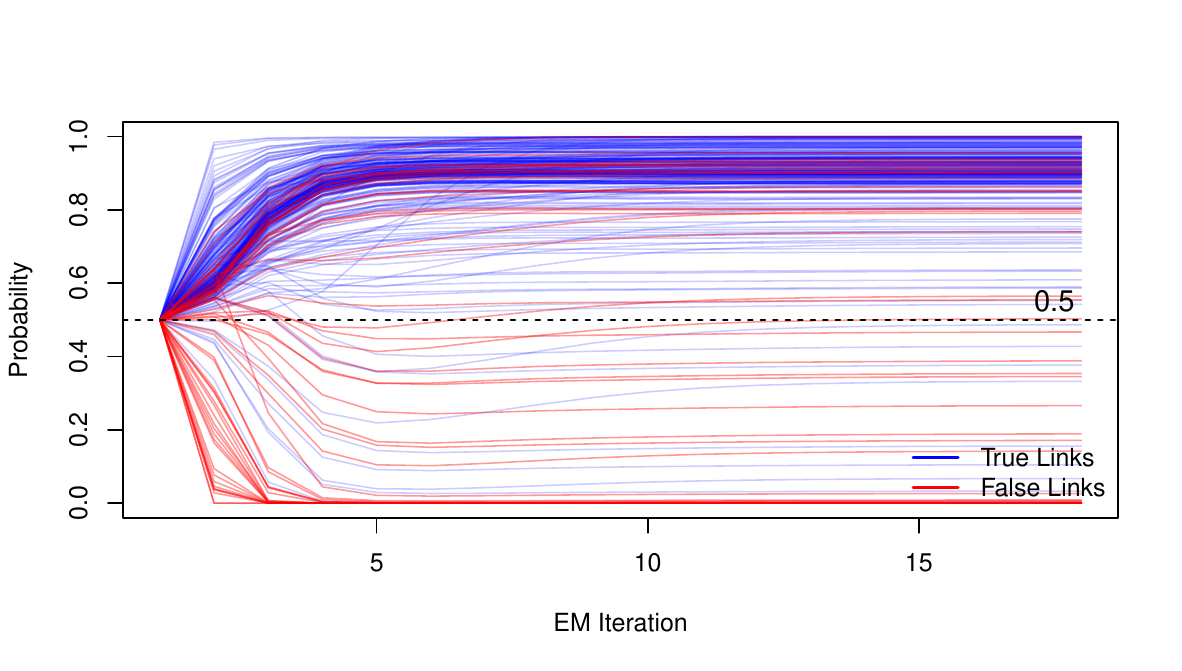}
    \caption{Latent class probabilities for all records for one $D^{(m)}$ using PLMIc. Results displayed over iterations of the EM algorithm.  
    }
    \label{fig:EM_single_simulation}
\end{figure}

Figure \ref{fig:EM_single_simulation} illustrates the latent class probabilities for each record in a single $D^{(m)}$ across iterations of the EM algorithm for PLMIc. The EM algorithm exhibits discriminatory power: false links (red lines) generally have probabilities below one and many approach zero, whereas the true links (blue lines) generally have probabilities near one. This separation occurs because false links create low density concatenations of $\boldsymbol x_k$ and $y_k$ that the algorithm can detect and filter.

\subsection{Robustness to Violations of Covariate-Linkage Independence}\label{sec:simu_violation_indep}

Both PLMIc and PLMI presume several conditional independence assumptions.  It is possible that they may  not hold in certain scenarios. In particular,  $\boldsymbol{x}$ could exhibit systematic differences across links and non-links, so that neither  A3 nor A4 are reasonable. In this section, we examine the performance of PLMIc and PLMI under this violation of the assumptions.

For records in $D_2$ that jave a true link in $D_1$, we set $x \sim \mathcal{N}(0,1)$. For records that do not have a true link, we set $x \sim \mathcal{N}(r,1)$, where $r \in \{1, 3, 5\}$. The different values of $r$ represent scenarios ranging from highly overlapping distributions to almost completely separable distributions. We generate  $\boldsymbol{Y}$ in $D_2$ conditional on $\boldsymbol X$ using a regression model with $R^2=0.3$.  We assume low error rates and 50\% overlap.  We use 5\% seeds for all methods.

\begin{table}[t]
\centering
\caption{Simulation results when generating $\boldsymbol X$ from a $N(0,1)$ for true links and from a $N(r,1)$ for false links, where $r \in (1,3,5)$. Results for 50\% overlap with low error rates and $R^2=0.3$, using $5\%$ of records as seeds. Results based on 500 simulation runs.}
\label{tab:Simu_Indep_Violation}
\scriptsize
\setlength{\tabcolsep}{2.5pt} 
\resizebox{\textwidth}{!}{
\begin{tabular}{c c *{12}{S[table-format=3.1]}}
\toprule
\textbf{Parameter} & \textbf{$r$} 
& \multicolumn{4}{c}{\textbf{Coverage} (\%)} 
& \multicolumn{4}{c}{\textbf{Median Absolute Diff.} ($\times 100$)} 
& \multicolumn{4}{c}{\textbf{Median Length} ($\times 100$)} \\ 
\cmidrule(lr){3-6} \cmidrule(lr){7-10} \cmidrule(lr){11-14}
& & {PLMIc} & {PLMI} & {TS.OLS} & {Perfect} 
& {PLMIc} & {PLMI} & {TS.OLS} & {Perfect} 
& {PLMIc} & {PLMI} & {TS.OLS} & {Perfect} \\ 
\midrule
\multirow{3}{*}{$\beta_1$} 
& 1 & 81.0 & 82.8 & 3.8 & 90.2 & 37.8 & 39.2 & 142.0 & 20.9 & 141.1 & 153.0 & 132.7 & 95.5 \\
& 3 & 83.6 & 83.4 & 3.6 & 90.4 & 35.6 & 35.3 & 139.5 & 19.5 & 140.2 & 148.7 & 133.9 & 95.7 \\
& 5 & 82.4 & 81.9 & 2.8 & 89.4 & 36.7 & 38.1 & 144.4 & 20.4 & 144.3 & 158.8 & 130.1 & 95.9 \\ 
\midrule
\multirow{3}{*}{$\beta_0$} 
& 1 & 88.6 & 88.8 & 81.6 & 91.2 & 20.5 & 20.7 & 27.9 & 10.0 & 102.4 & 102.5 & 106.5 & 95.7 \\
& 3 & 86.8 & 86.5 & 80.8 & 90.0 & 21.7 & 21.7 & 27.4 & 19.1 & 102.2 & 101.7 & 106.3 & 95.8 \\
& 5 & 86.6 & 86.9 & 83.0 & 88.4 & 21.4 & 21.8 & 29.4 & 18.5 & 102.6 & 102.4 & 105.9 & 95.3 \\ 
\bottomrule
\end{tabular}
}
\end{table}

Table~\ref{tab:Simu_Indep_Violation} displays the results. Intuitively, the violation of independence increases the cost of declaring false links. This cost becomes higher as the covariate distributions for links and non-links separate.
TS.OLS deteriorates: coverage rates for $\beta_1$ drop to as low as 2.8\%, accompanied with large absolute differences. In contrast,  PLMIc and PLMI models demonstrate  stability across the values of $r$. For $\beta_1$, coverage rates are reduced but remain consistently between 81\% to 84\% across all scenarios. 

\subsection{Extension to Multiple Linear Regression}\label{sec:simu_multiReg}

PLMIc and PLMI can be readily extended to multiple regression with covariates that are not normally distributed. The key change is that the marginal distribution of $\boldsymbol Y$ is no longer normally distributed and needs to be estimated, e.g., with a kernel density estimator.

To illustrate, we consider the regression model $y = \beta_0 + \beta_1 x_1 + \beta_2 x_2 + \varepsilon$ where $\varepsilon \sim \mathcal{N}(0,1)$. We simulate $x_1$ from $\mathcal{N}(0,1)$ and $x_2\sim \text{Bernoulli}(0.5)$. We set $\beta_0=3$, $\beta_1=0.5$, and $\beta_2 \in (1, 10)$, so that we have $R^2=0.33$ and $R^2=0.96$. 
The marginal distribution of $\boldsymbol Y$ is a mixture of two normal distributions centered at different means. 
We estimate this distribution by simply using the empirical distribution of $y$ for all records in $D_2$.
For the linkage scenario, we presume 250 overlap, low error rates, and 5\% seeds.  We generate 500 simulation runs from this scenario.

Table~\ref{tab:Simu_Beta_x2} summarizes the results.  PLMIc and PLMI continue to perform well with close to 90\% nominal coverage rates.  The median absolute differences in the point estimates and average interval lengths are close to those for Perfect. Evidently, these models can be effective with multiple repression and do not require normally distribution covariates.  We note that TS.OLS is less effective. It has lower coverage for $\beta_2$ when $R^2=0.96$ and overly conservative coverage for $\beta_1$, with interval lengths that are twice as wide as those from Perfect.

\begin{table}[t]
\centering
\caption{Inference of $\beta_0$, $\beta_1$, and $\beta_2$ across different $beta_2 = 1, 10$, with $n_{\text{error}}=1$ and 5\% seeds. PLMIc and PLMI are two proposed models, TS.OLS is the two-step OLS model, and Perfect is the OLS with only true links. A total of 500 simulations are performed for each scenario.}
\label{tab:Simu_Beta_x2}
\scriptsize
\setlength{\tabcolsep}{2.5pt}
\resizebox{\textwidth}{!}{
\begin{tabular}{ccrrrrrrrrrrrr}
\toprule
\textbf{Parameter}         & \textbf{$R^2$} & \multicolumn{4}{c}{\textbf{Coverage} (\%)} & \multicolumn{4}{c}{\textbf{Median Absolute Diff.} ($\times 100$)} & \multicolumn{4}{c}{\textbf{Median Length} ($\times 100$)} \\ \cmidrule(lr){3-6} \cmidrule(lr){7-10} \cmidrule(lr){11-14} 
                           &                & PLMIc    & PLMI    & TS.OLS    & Perfect   & PLMIc          & PLMI         & TS.OLS         & Perfect         & PLMIc        & PLMI        & TS.OLS       & Perfect       \\ \midrule
\multirow{2}{*}{$\beta_2$} & 0.33           & 90.0       & 90.0      & 92.0        & 89.0        & 8.0            & 8.9          & 9.2            & 8.2             & 42.4         & 41.9        & 42.8         & 41.7          \\
                           & 0.96           & 89.0       & 87.0      & 66.0        & 89.0        & 8.6            & 8.3          & 38.0            & 8.8             & 41.6         & 40.4        & 90.3         & 41.8          \\ \midrule
\multirow{2}{*}{$\beta_1$} & 0.33           & 88.2     & 87.6    & 88.8      & 89.6      & 43.8           & 43.4         & 43.0           & 43.9            & 21.4         & 21.1        & 21.6         & 20.9          \\
                           & 0.96           & 91.6     & 91.0    & 98.2      & 92.0      & 41.9           & 41.7         & 51.9           & 43.1            & 20.8         & 20.2        & 41.7         & 20.9          \\ \midrule
\multirow{2}{*}{$\beta_0$} & 0.33           & 89.0     & 89.0    & 89.4      & 88.0      & 7.3            & 7.3          & 7.3            & 7.5             & 30.0         & 29.7        & 30.3         & 29.6          \\
                           & 0.96           & 88.0     & 86.8    & 85.2      & 86.6      & 7.2            & 7.2          & 20.1           & 7.3             & 29.4         & 28.5        & 61.1         & 29.5          \\ \bottomrule

\end{tabular}
}
\end{table}

\section{Illustration Using Genuine Data}

In this section, we construct a record linkage scenario using data  from the Bank of Italy's Survey on Household Income and Wealth (SHIW). These data were used by \textcite{guha2022bayesian} to evaluate fully Bayesian inference for simultaneous causal inference and record linkage, which is a different methodology than considered here.

\subsection{Data Description and Background}
The SHIW is a biennial, nationally representative survey conducted since 1965 (with 1998 replacing the originally scheduled 1997 wave).
 The survey captures information about Italian families' economic and financial circumstances, including household demographics, expenditure patterns, and financial instrument usage. We use data from the 1995 and 1998 waves with households as our unit of analysis. These data include identifiers so that longitudinal histories are available. However, for our constructed record linkage, we take a subset of the 1995 data as $D_1$ and a subset of the 1998 data as $D_2$ to construct a record linkage scenario.

We use gender and education of household head as linking variables. We also introduce linking variables by borrowing the names and birth dates (year, month, day) of records in the \texttt{RL10000} dataset in \texttt{R} library \texttt{RecordLinkage}. The linking variables in the \texttt{RL10000} data are designed to have realistic errors for records that are matches. To construct our simulated data, we assign linking variables as follows: for individuals who appear in both waves (true matches), we use the two different records of the same individual from \texttt{RL10000}. For individuals who appear in only one wave (non-matches), we randomly sample from the 8,000 unique individual records. This approach ensures that true matches have correlated errors in their linking variables across waves, while non-matches have independent linking information.

\begin{table}[t]
\centering
\caption{Mapping rules for comparison vector $\boldsymbol{\gamma}$ in household income case study.}
\label{tab:comp_vec_realdata}
\begin{tabular}{llccccc}
\toprule
\textbf{Fields}  & \textbf{Similarity}  & \multicolumn{5}{c}{\textbf{Levels of agreement}} \\
\cmidrule(l){3-7}
 &  & 1 & 2 & 3 & 4 & 5\\
\midrule
First name      & Levenshtein & 0    & (0, 0.001]        & (0, 0.25]     & (0.25, 0.5]    & (0.5, 1] \\
Last name       & Levenshtein & 0    & (0, 0.001]        & (0, 0.25]     & (0.25, 0.5]    & (0.5, 1] \\
Year      & Exact match & 0            & 1             & —              & —    & —    \\
Month     & Exact match & 0            & 1             & —              & —    & —    \\
Day       & Exact match & 0            & 1             & —              & —    & —    \\
Gender    & Exact match & 0            & 1             & —              & —    & —    \\
Education & Exact match & 0            & 1             & —              & —    & —    \\
\bottomrule
\end{tabular}
\end{table}

To construct $D_1$ and $D_2$
we randomly sample one subset from each of the 1995 and 1998 files. Thus, $D_2$ include  $n_2=1582$ households from the 1998 data and $D_1$ includes $n_1=1591$ households from the 1995 data.
Approximately 50\% of the records (760 pairs) in $D_1 \cup D_2$ represent true matches, while the remaining records are unique to each year.  We compute the comparison vector,  $\boldsymbol{\gamma}$, using the seven linking variables based on the rules in Table~\ref{tab:comp_vec_realdata}.

As the response variable $\boldsymbol{Y}$, we use net disposable income measured in 1998, which is available only in $D_2$. As the covariates $\boldsymbol X$, we use variables from the 1995 survey including real net disposable income (lagged response variable, $\boldsymbol{X}_1$), education level ($\boldsymbol{X}_2$, indicator of $1$ for at least middle-school and 0 otherwise), gender ($\boldsymbol{X}_3$, indicator of $1$ for male and 0 otherwise), and cash holdings ($\boldsymbol{X}_4$).  We remove observations in the top and bottom 5\% of $\boldsymbol{Y}$ and $\boldsymbol{X}_1$ to mitigate outlier effects, and standardize  $\boldsymbol{Y}$, $\boldsymbol{X}_1$, and $\boldsymbol{X}_4$ before record linkage. We randomly sampled 50 seeds from the set of records appearing in both datasets to help with inference. 

\subsection{Results}
We implement both the PLMI and PLMIc models, along with a two-stage approach and the analysis using the true links only. We again refer to the latter two approaches as TS.OLS and Perfect. We use the same prior distributions as in the simulation studies. We estimate the marginal density $p_{Y}$ using the empirical distribution of all records in $D_2$. We run MCMC chains for 1000 iterations, discarding the first 500 as burn-in, and base inferences on the remaining 500 iterations. Diagnostic checks offer no evidence that the samplers fail to converge.

We consider two sets of linking variables. In the first, we use all but names for linkages. This obviously does not offer strong information to distinguish linked and nonlinked pairs. In the second, we use all seven linking variables. 
The name variables have substantial discriminatory power. When name fields are included, the Bayesian record linkage model estimates an average of 775 links a posteriori, of which 758 are true matches for an accuracy rate of 97\%. When the name variables are not used, the Bayesian record linkage model estimates an average of 780 links, with only 465 as true links for an accuracy rate of 59\%. 

\begin{table}[t]
\centering
\small
\setlength{\tabcolsep}{3pt} 

\caption{Point estimates and 90\% confidence intervals for the regression coefficients in the constructed SHIW record linkage illustration.}
\label{tab:household_inference}

\begin{tabular}{
l
r r
r r
r r
r r
r r
}
\toprule
        & \multicolumn{2}{c}{$\beta_0$}
        & \multicolumn{2}{c}{$\beta_1$}
        & \multicolumn{2}{c}{$\beta_2$}
        & \multicolumn{2}{c}{$\beta_3$}
        & \multicolumn{2}{c}{$\beta_4$}
        \\
\cmidrule(lr){2-3} \cmidrule(lr){4-5} \cmidrule(lr){6-7} \cmidrule(lr){8-9} \cmidrule(lr){10-11}
Model   & {Est.} & {90\% CI}
        & {Est.} & {90\% CI}
        & {Est.} & {90\% CI}
        & {Est.} & {90\% CI}
        & {Est.} & {90\% CI}
        \\
\midrule

Perfect
& -0.18 & {[-0.31, -0.06]}
&  0.70 & {[ 0.65,  0.74]}
&  0.14 & {[ 0.05,  0.23]}
&  0.20 & {[ 0.07,  0.33]}
& -0.01 & {[-0.06,  0.03]}
\\
\midrule

\multicolumn{11}{l}{\textbf{Panel A}: without name fields} \\
\midrule

TS.OLS
& -0.14 & {[-0.35,  0.08]}
&  0.26 & {[ 0.19,  0.34]}
&  0.06 & {[-0.08,  0.21]}
&  0.11 & {[-0.10,  0.33]}
& -0.04 & {[-0.11,  0.02]}
\\

PLMIc   
& -0.26 & {[-0.43, -0.09]}
&  0.71 & {[ 0.62,  0.79]}
&  0.12 & {[-0.04,  0.28]}
&  0.22 & {[ 0.05,  0.39]}
& -0.06 & {[-0.12, -0.00]}
\\

PLMI    
& -0.26 & {[-0.44, -0.07]}
&  0.69 & {[ 0.20,  1.19]}
&  0.12 & {[-0.06,  0.30]}
&  0.22 & {[-0.01,  0.44]}
& -0.06 & {[-0.18,  0.05]}
\\

\midrule
\multicolumn{11}{l}{\textbf{Panel B}: with name fields} \\
\midrule

TS.OLS
& -0.19 & {[-0.32, -0.06]}
&  0.67 & {[ 0.62,  0.72]}
&  0.12 & {[ 0.02,  0.22]}
&  0.20 & {[ 0.07,  0.33]}
& -0.02 & {[-0.06,  0.03]}
\\

PLMIc   
& -0.20 & {[-0.31, -0.09]}
&  0.76 & {[ 0.72,  0.80]}
&  0.16 & {[ 0.07,  0.24]}
&  0.19 & {[ 0.08,  0.31]}
& -0.04 & {[-0.08, -0.00]}
\\

PLMI    
& -0.21 & {[-0.32, -0.09]}
&  0.76 & {[ 0.72,  0.80]}
&  0.15 & {[ 0.06,  0.24]}
&  0.19 & {[ 0.07,  0.31]}
& -0.04 & {[-0.08,  0.00]}
\\

\bottomrule
\end{tabular}
\end{table}

Table~\ref{tab:household_inference} displays the estimates for the coefficients. In the Perfect model, the estimates of these coefficients have $t$-statistics of -2, 26, 2, 2 and -0.5, respectively. Without using the name fields, the point estimates for PLMIc and PLMI are closer to those from the Perfect model benchmarks than those from TS.OLS. 
The TS.OLS estimates are biased toward zero due to the high proportion of false links, whereas PLMI and PLMIc downweight incorrect linkages in the estimation, which reduces bias. The interval width of PLMI is in general wider than PLMIc, suggesting that linkage quality is helpful for filtering true links and reducing the likelihood of the EM algorithm converging to a local optimum. 
When using the name fields, all models perform nearly as well as Perfect due to the high quality of the linking fields. Taking both sets together, we conclude that PLMIc and PLMI improves estimation when needed and does not harm estimation when linkage quality is high. 

The covariate $\boldsymbol X_1$ provides a particularly strong signal, with a $t$-value of 26 in the Perfect model. PLMIc and PLMI leverage this signal strength when seeking to identify correct linkages. The improved linkage quality subsequently enhances the estimation accuracy of other coefficients as well.

\section{Discussion}\label{sec:discussion}

The findings from the simulations highlight the potential advantages of using PLMIc and PLMI for parameter inference in  regression with linked data. These methods frequently outperform the TS.OLS approach--especially when $R^2$ is high---often achieving coverage rates and median absolute differences in point estimates comparable to those from using only the true links. We also observe that PLMIc and PLMI can improve dramatically with even a small number of seeds.  This finding about seeds
suggests that it may be worthwhile to invest in clerical review to seeds.

Of course, PLMIc and PLMI have limitations. First, they presume a reasonable model specification for the regression.  If the regression is poorly specified, PLMIc and PLMI will not yield reliable results. Analysts should use any available seeds for exploratory data analysis and check residual diagnostics, for example, computing the diagnostics in several representative $D^{(m)}$ for records that are highly likely to be true links, e.g., have high latent class probabilities.
Second, because the methods start with plausibly linked datasets, they only can downweight false links.  They are not able to declare new links from the records in $D_1\cup D_2$. Third, running an EM algorithm in each of $M$ datasets increases computational complexity.  Fortunately, given $M$ plausibly linked datasets, one can easily run the EM algorithms in parallel.

There are opportunities for future extensions of the PLMIc and PLMI models. 
In particular, the EM algorithms are estimated within each plausible linked dataset. It may be possible to pool information across $D^{(1)}, \dots, D^{(M)}$ when estimating the latent class probabilities, which potentially could sharpen  inferences about each $t_{k}^{(m)}$.  This could increase computational overhead, however, and so any benefits for accuracy must be weighed against computational costs.
Additionally, the strategy used for PLMIc and PLMI can be adapted for other parametric models. One would need to tailor features of the model, such as the marginal distribution of the outcomes, to the model at hand.

\printbibliography
\end{document}